\documentstyle[aps,eqsecnum,prbbib]{revtex}
\begin{document}
%\draft
\title{Nonlinear electrodynamics of p-wave superconductors}
\author{Klaus Halterman\cite{klaus}  and Oriol T. Valls\cite{oriol}}
\address{School of Physics and Astronomy and Minnesota Supercomputer Institute
\\ University of Minnesota \\
Minneapolis, Minnesota 55455-0149}
\date{\today}
\maketitle
\begin{abstract}
We consider the Maxwell-London electrodynamics of 
three dimensional superconductors in p-wave 
pairing states with nodal points or lines in the energy gap. 
The current-velocity 
relation is then nonlinear in the applied field, cubic for point
nodes 
and quadratic for lines.
We obtain explicit angular and depth dependent 
expressions for measurable quantities such as the transverse magnetic moment,
and associated torque. These dependences are different
 for point and line nodes and can be used
to distinguish between different order parameters. 
We discuss the experimental feasibility of this method, and bring forth 
its advantages, as well as limitations that might be present.
\end{abstract}

% body of paper here
\section{Introduction}

The number and variety of superconducting materials for which evidence
of exotic Cooper pairing (i.e., pairing in a state other than the
usual s-wave) exists is constantly increasing. For high temperature 
superconducting oxides (HTSC's) the consensus\cite{htsc} is indeed that 
the pairing
state is, in nearly all cases, at least predominantly d-wave, specifically
of the $d_{x^2-y^2}$ form, with lines of nodes. 
Rather persuasive (although not conclusive)
evidence in the form of both
experiments and theoretical arguments,\cite{ruth,ruth2,ruth4} 
has recently been brought forward for
p-wave superconductivity in ${\rm Sr_2RuO_4}$. The pairing state
currently favored by many
is of the same form\cite{sameform,sameform2,maki,rice} as
that of the A phase\cite{h3} in ${\rm ^3He}$, which has
point nodes.
Several heavy fermion (HF) materials, the discovery of 
which predated that of HTSC's but for which determinations of the pairing state
have proved harder to achieve, are now also 
believed with varying degrees of certainty, to belong
in the exotic camp\cite{hf,hf2,hff3}.
There are also results indicating\cite{kan96,organic2}
that superconducting families of organic
salts such as ${\rm \kappa-(BEDT-TTF)_2C u (NCS)_2}$ and 
 ${\rm (TMTSF)_2X}$ 
(${\rm X = PF_6,ClO_4}$, etc.) 
also exhibit 
unconventional superconductivity. In some cases it has been argued that
the  pairing appears to be in the p-wave\cite{torganic1,organic1}. 

Determination of pairing states is not easy, particularly if one wishes
to know more details than merely their overall symmetry. Even
in the best studied HTSC's, questions such as what is
the angle between lines of nodes in orthorhombic
compounds, or whether true nodes,
rather than very deep minima, exist, are still matters for occasionally
heated debate. The situation is much worse for the other materials
mentioned, 
where the evidence is much more
preliminary, and at times contradictory.\cite{spec,counter,mayo,miya,lang,hase}
The determination of
the pairing state is often hampered by difficulties in interpreting results.
Regions (points or lines) where the energy gap vanishes are often 
the signature of exotic pairing (but
not invariably, the B phase of ${\rm ^3He}$ is a well-known counterexample).
These ``gap nodes'' lead to various power law behaviors for quantities
that otherwise would behave exponentially with temperature, but sometimes
there are alternative explanations for the power laws.
It is difficult moreover, to distinguish  between experimental outcomes
arising from zeroes in the energy gap and those arising only from strong
anisotropy. An additional complication is that for non s-wave
superconducting materials, the order
parameter (OP) state at the surface may easily differ\cite{safi} 
from that in the bulk.

It is therefore important to study probes of the OP 
symmetry able to discern as unambiguously as possible details of
the pairing state, such
as the existence, nature  and position of the {\it bulk} OP nodes.
One such probe is afforded by the nonlinear Maxwell-London electrodynamics
of exotic pairing states
in the Meissner regime. Electrodynamic effects probe the sample over
a scale determined by the penetration depth $\lambda$, which is large for
the materials of interest. It was pointed out\cite{ys} in the context
of d-wave superconductivity that order parameter nodes lead to observable
nonlinear effects at low temperatures, the chief quantities
of experimental interest being the magnetic field dependent
penetration depth $\lambda(H)$, the 
nonlinear transverse component of the magnetic
moment, $m_\perp$, induced by the application
of a magnetic field, and the torque associated with this 
transverse moment.\cite{sv,ys1,zv1}. 
Further developments of the method, always in the context of 
predominantly d-wave
superconductivity, showed\cite{zv2} that it can be used to
perform {\it node spectroscopy}, that is, to infer in detail the angular
structure of the regions where the order parameter vanishes (nodes) or
is very small (``quasinodes'').

These developments took place within the study of the high temperature oxide
superconductors. For these materials, the temperature scales as set
by $T_c$ are higher and achieving the required low temperature conditions
is very easy. However, recent improvements in experimental
techniques involving torsional oscillators\cite{to} and torque
magnetometry\cite{tm} make it possible
to measure extremely small 
moments and torques at dilution refrigerator temperatures. 
Experiments to accurately measure $\lambda$ in that temperature range
are also\cite{allen} being planned. Thus, the relevant region for
performing nonlinear electrodynamics experiments in low $T_c$ materials
is becoming accessible.

With this in mind,  we take up 
in this work the question of the use of
methods based on nonlinear electrodynamics to study  
exotic superconducting materials, other than HTSC's. Specifically, we 
will consider here simple OP's 
both with point nodes and with three-dimensional nodal lines, as would
occur for example in p-wave superconductivity. Our
efforts will focus on the calculation of the dependence
of $m_\perp$ (or its associated torque) on the 
magnetic field and the appropriate angle of
rotation.  We also compute the field dependence of the low temperature
penetration depth. 
We will present 
estimates based on published values of the relevant material parameters 
showing that the required measurements appear to be technically feasible. 
These estimates are presented, not to prejudge the pairing state associated 
with any 
material, but rather to show the expected signal 
{\it if}
 the material indeed does 
have the assumed OP. 

In the next Section we  introduce the geometries and the order parameter
forms that we study. We then calculate the nonlinear
relation between current and superfluid flow field, using 
an extension of the three
dimensional methods of Ref.\onlinecite{zv1}. From these relations,
we obtain the physical quantities
of interest, through the appropriate generalization of existing\cite{zv2}
perturbation methods. In Section \ref{disc} we summarize our results,
and consider the question of the
experimental feasibility of  using this method
on several materials. We conclude with a discussion of the  advantages
and limitations of the method and of the specific treatment presented
in this work.

\section{Methods and Results}
\label{methods}

\subsection{Maxwell-London Electrodynamics}
We first briefly outline the nonlinear Maxwell-London equations, on which
our method is built. We will
not dwell into any details that were discussed
elsewhere.\cite{ys,sv,zv1}.
When a magnetic field ${\bf H}_a$ is applied to a superconductor a
superfluid flow  field ${\bf v(r)}$ is set up.
The relation between  ${\bf v(r)}$ and the local magnetic field 
${\bf H(r)}$ is given\cite{below} by the
second London equation:
\begin{equation}
\nabla \times {\bf v}=\frac{e}{c}{\bf H}.
\label{londoneq}
\end{equation}
where $e$ is the proton charge.
Amp\`ere's law for steady-state currents, $\nabla\times 
{\bf H}=\frac {4 \pi}{c}{\bf j}$, 
can be combined with Eq. (\ref{londoneq}) to obtain:
\begin{equation}
\nabla\times\nabla\times{\bf v}=\frac{4 \pi e}{c^2}{\bf j(v)}.
\label{maxlon}
\end{equation}

In this equation there are still two unknown fields.
For a solution to be obtained, the
functional relationship between ${\bf j}$ and ${\bf v}$ is needed. This 
can be found\cite{bard} by using the  two-fluid model. The 
quasiparticle excitation spectrum,
$E(\epsilon)=(\epsilon^2+\left| \Delta(s) \right|^2)^{1/2}$,  is 
modified by a Doppler shift to $E(\epsilon)+{\bf v}_f\cdot {\bf v}$. 
Here 
${\epsilon}$ is the quasiparticle
energy referred to the Fermi surface,
$\Delta(s)$ denotes the OP 
dependence 
on the point $s$ on the Fermi 
surface\cite{par}, and
${\bf v}_f$ is the Fermi velocity.
This leads to a relation between ${\bf j}$ and ${\bf v}$ 
of the form
\begin{equation}
{\bf j}({\bf v})= {\bf j}_{lin}({\bf v})+{\bf j}_{nl}({\bf v}).
\label{jofv}
\end{equation}
After some algebra\cite{ys,zv1}, the linear and nonlinear parts can be
written, respectively, as:
\begin{mathletters}
\label{nonlinjv}
%\begin{eqnarray}
\begin{equation}
{\bf j}_{lin}({\bf v}) = -eN_f\int_{FS} d^2s \: n(s)
 {\bf v}_f ({\bf v}_f \cdot {\bf v}), 
 \label{nonlina}
 \end{equation}
 \begin{equation}
{\bf j}_{nl}({\bf v}) = -2eN_f\int_{FS} d^2s \: n(s) {\bf v}_f 
\int^{\infty}_0 d\epsilon \: f(E(\epsilon)+{\bf v}_f\cdot {\bf v}),
\label{nonlinb}
\end{equation}
%\end{eqnarray}
\end{mathletters}
where $N_f$ is the total density of states at the Fermi level,
$n(s)$ the local density of states at the Fermi surface (FS),
normalized to unity,
and $f$ the Fermi function.
The first term in (\ref{jofv}), given
by (\ref{nonlina}), is the usual linear relation
${\bf j}_{lin}=-e \tilde{\rho} {\bf v}$, where $\tilde{\rho}$ is the superfluid
density tensor. At $T=0$
the nonlinear (in ${\bf v}$) corrections,  described by 
(\ref{nonlinb}), can be 
written as:
\begin{equation}
{\bf j}_{nl}{\bf (v)}= -2 eN_f\int_{FS} d^2s \: n(s) {\bf v}_f  
\Theta(-{\bf v}_{f} \cdot {\bf v}-\left| \Delta(s) \right| ) 
[({\bf v}_{f} \cdot {\bf v})^2-\left| \Delta(s) \right|^2]^{1/2} 
\label{jq},
\end{equation}
which is valid for any $\Delta(s)$. 
The key point is that the step 
function in Eq.(\ref{jq}) restricts the integration over the FS by 
\begin{equation}
\left| \Delta(s) \right|  +{\bf v}_f \cdot {\bf v} <0.
\label{res}
\end{equation}
Thus, when the OP has nodes (or very deep minima) in the Fermi surface,
only regions near these nodes participate in 
populating the quasiparticle spectrum. 
The integration is dominated by contributions from these
regions, and can be written
as\cite{zv2} a sum over local contributions from each of them.
The values of the Fermi velocity in the integrand can be replaced by their {\it local}
values at the corresponding nodal region.
We need, therefore, more information on the 
geometry of the superconductor and the angular dependence of the 
order parameter 
to carry out the above integration. 

\subsection{Geometry and order parameters}

We will consider superconducting materials of orthorhombic or higher
symmetry and denote the crystallographic axes as $a$, $b$ and $c$, corresponding
as customary to the $x$, $y$ and $z$ directions. We will assume that the
samples are infinite in a plane parallel to the direction
of the applied field, and of thickness $d$ in the direction
normal to this plane. This allows us to solve (\ref{maxlon}) and (\ref{jofv})
analytically. Effects of the sample finite extension in the plane would
have to be taken into account numerically, but this is unnecessary
since it has been shown in the context of d-wave 
pairing\cite{zvcomp}, that such effects merely lead to a small increase
in the amplitude of the nonlinear signal, and to no change in its angular
or field dependence. On the other hand, the effects of the thickness $d$ 
are very
important and we will include them fully.

We will consider two simple types of p-wave OP's in this paper.
The first is representative of the case where the OP has point nodes,
as might occur \cite{ruth2,ruth4,sameform2}
in ${\rm Sr_2RuO_4}$. 
Up to a phase factor (the nonlinear electrodynamic
effects depend only on the absolute value of the OP) we write 
for the angular dependence of the OP near the nodes:
\begin{equation}
\Delta(\theta)=\Delta_0 \sin(\theta),
\label{deltasin}
\end{equation}
where $\theta$ is a polar angle. Only the local
properties at the nodes are important: the form
(\ref{deltasin}) is
assumed only near the nodes, e.g. 
near $\theta=0$ it means $\Delta(\theta)\approx \Delta_0\theta$. 
Thus, the parameter $\Delta_0$ must be thought of as the slope of
the OP near the node, rather than its maximum value.
The second type we will consider
is a prototype of  OP's with line nodes, as they might
occur in  some heavy 
fermion compounds\cite{hf2} or even in\cite{hase} ${\rm Sr_2RuO_4}$.
Again, up to an unimportant
phase factor, we assume the form for the angular dependence 
near the nodal line:
\begin{equation}
\Delta(\theta)=\Delta_0 \cos(\theta),
\label{deltacos}
\end{equation}
where the above warning as to the interpretation of $\Delta_0$ as the slope
near the nodal region must be repeated.
These two forms are archetypes for the possibly more intricate forms of
the angular dependence of the OP in real materials. In non s-wave superconductors,
 the OP need not belong to a one-dimensional
representation. Because of this,
there may be OP collective modes\cite{maki,tewordt} and
internal structure effects that are not included
in our considerations. The angular dependence of the OP, however, will
be in a solid very strongly pinned\cite{maki}
by crystal effects (this is obviously
not the case in liquid ${\rm ^3He}$). The internal structure of the OP 
should then not affect our
nonlinear results, since only the application of small dc fields
is involved.

 The experimental setup we envision would involve applying a field
 parallel to the $a-c$ plane, with the direction normal to the
 slab being along the $b$ axis\cite{rotate}. The 
sample would then be rotated about the $b$ axis while the magnetic field 
remains fixed. The currents then flow in various directions 
depending on their orientation relative to the OP under consideration.
We will denote by $\psi$ the angle between the applied field ${\bf H}_a$ 
and the $z$ axis and we will investigate the angular dependence of the 
transverse magnetic moment or the torque as a function of $\psi$. We will also 
calculate the field dependence of the penetration depth for the directions
of symmetry.

In this geometry we can solve the problem
analytically. 
When ${\bf H}_a$ is applied in the $a-c$ plane, 
the fields have only $x$ and $z$ components, which depend   
 {\it only} on the coordinate $y$.  Eq. (\ref{maxlon}) then reduces to
\begin{equation}
\frac{d^2{\bf v}}{d y^2}+\frac{4 \pi e}{c^2}{\bf j(v)}=0.
\label{lap}
\end{equation}
For our given geometry, $j_{nl i}$
and $v_i$ have odd parity with respect to the $y$ 
coordinate, so it is sufficient to solve the boundary value problem 
for $y \geq 0$.                                                    
The two required boundary conditions are:
\begin{mathletters}
\begin{eqnarray}
\label{bc}
\frac{c}{e}\left.(\nabla \times {\bf v}) \right |_{y=d/2}&=& {\bf H}_a , \\
{\bf v}|_{y=0}&=&0.
\end{eqnarray}
\end{mathletters}
We can now proceed
to explicitly calculating 
the nonlinear currents.

\subsection{Nonlinear currents}
First, we carry out the integration in Eq. (\ref{jq}). This
can be performed exactly
on a three dimensional Fermi surface, without the need to take recourse
to the approximations discussed in Appendix A of Ref.\onlinecite{zv1}.
The relevant regions of integration as discussed below
(\ref{res}) are contained 
within a small range near the nodes, with boundaries
that can be expressed in terms of limiting angles
$\theta_c$, as determined from
$({\bf v}_{f} \cdot {\bf v})^2=\left| \Delta(\theta_{c}) \right|^2$.

Consider first the OP given in Eq.(\ref{deltasin}). In this case
${\bf v}_f(s)$ can be replaced in the relevant regions of the
integrand by its local value, along the $z$ axis, at the nodes.
By symmetry, we can restrict ourselves to the node at $\theta=0$
since the contribution from $\theta=\pi$ is identical.
Thus, we have  ${\bf v}_f \approx (0,0,v_{fz} )$, and 
the restriction (\ref{res}) means that 
in performing the integral in (\ref{jq}), we can
replace $\int_{FS} d^2s \: n(s)$ by 
$\int_{\Omega_c}  \: d\phi  \theta 
d\theta/4 \pi$, where $\Omega_c$ denotes
the region $|\theta|<\theta_c$, with
$\theta^2_{c}=(v_{fz} v_z)^2/{\Delta_{0}^2}$,
and
with no 
restrictions on $\phi$. 
It
is easy to see that this yields 
only a $z$-component to the nonlinear current. The integrals are 
elementary and one finds:
\begin{equation}
j_{nlz}=\frac{1}{3}e N_f \frac{v_{fz}^4}{\Delta_{0}^2}v_{z}^3,
\label{jsin}
\end{equation}
where $\Delta_0$ the local
gap slope.
              
For the case of an order parameter as given
in (\ref{deltacos}), where the nodal line is at $\theta = 
\pi/2$, we can take $v_{fz}=0$ over the region of integration.
which is then limited to $|\theta-\pi/2|<\theta_c$,
where
$(\theta_c-\pi/2)^2=\left(\cos \alpha v_f v_{\perp}/\Delta_0\right)^2$. 
Here $v_{\perp}$ is the projection of ${\bf v}$ on the $x-y$ plane, and 
$\alpha$ the angle between $v_{\perp}$ and the in-plane $v_f$.  We make the 
replacement 
$\int_{FS} d^2s \: n(s)\rightarrow \int_{\Omega_c}  
\:  d\phi d\theta/4 \pi$, where $\Omega_c$ is the
region of integration as defined by $\theta_c$.
For
orthorhombic symmetry, $v_f$ depends on $\phi$.
We then transform the
integral over $\phi$ to one over $\alpha$ using the relation 
$\phi = \beta + \alpha$, where $\beta$ is the (fixed) angle
$v_{\perp}$ makes with the $x$ axis, and $\alpha$ is restricted
(from    
Eq. 
(\ref{res})) to $\pi/2<\alpha<3\pi/2$. The integration 
is lengthier but straightforward and results in two 
components to the current. The $x$ component is:
\begin{equation}
j_{nlx}=\frac{1}{3}e N_f 
\frac{v_{f x}^3}{\Delta_{0}}v_{x}\left(v_{x}^2+v_{y}^2\right)^{1/2}
{\Lambda}_x({\bf v}), 
\label{jcos}
\end{equation}
where ${\Lambda}_x({\bf v}) \equiv(1/5)[(3+ 2\delta^2) +(1-\delta^2)(v_x^2/(v_x^2+
v_y^2))]$ ($\Lambda_x \equiv 1$ when the material has tetragonal 
symmetry), $v_{fx}$ is the Fermi speed along the $x$ axis and 
the $a-b$ plane anisotropy is characterized by
$\delta \equiv \lambda_{x}/\lambda_{y}$, where $\lambda_{i}$ 
denotes the penetration depth along  the $i$ 
direction. The $y$ component, $j_{nly}$, is obtained by making
the obvious replacements in (\ref{jcos}). The coefficients in
(\ref{jsin}) and (\ref{jcos}) are given in terms of the local
values of the Fermi velocity and are therefore independent of the detailed shape
of the Fermi surface.

These expressions for ${\bf j}$ as a function of ${\bf v}$ can  be 
inserted into Eq. (\ref{lap}), which then
becomes a nonlinear differential equation in terms
of the flow field only. Implementing a perturbation scheme to lowest order in the 
flow field will allow exact expressions for both OP's to be obtained. This is 
addressed in the next subsection. 

\subsection{Perturbation solution}
For an OP of the form (\ref{deltasin}), we can insert Eq. (\ref{jsin}) 
into Eq. (\ref{lap}). We can then write the 
equation for the component carrying the nonlinear 
term as:
\begin{equation}
\frac{\partial^2 v_z}{\partial Y_z^2}-v_z
+\left(\frac{v_{z}}{v_{c z}}\right)^2v_{z}=0,
\label{usin}
\end{equation}
where we have introduced the dimensionless coordinate $Y_i\equiv y/\lambda_{i}$. 
The local critical velocity is defined as $v_{c i}\equiv 
{\Delta_{0}}/{v_{f i}}$, (for $i=x,y,z$) and we have used the
three dimensional relation 
${1}/{\lambda_i^{2}}=({4 \pi e^2}/{3c^2})N_f v_{fi}^2$.  From 
Eq. (\ref{bc}), the boundary condition at the surface $Y_i=Y_{i s} \equiv 
{d}/(2 \lambda_{i})$ can be written in terms of our new variables:
\begin{equation}
\label{bcds}
\left. \frac{\partial v_{x}}{\partial Y_x} \right 
|_{Y= Y_{x s}}=-\frac{{e}{\lambda_{x}}}{c}H_a\cos \psi, \qquad 
\left. \frac{\partial v_{z}}{\partial Y_z} \right 
|_{Y_z=Y_{z s}}=\frac{{e}{\lambda_{z}}}{c}H_a\sin\psi. 
\end{equation}
We now expand $v_z(Y_z)$ to first order in the 
parameter 
$\alpha_z=({v_{f z}}/{\Delta_0})^2$,
which is small in the
 typical experimental situations. We write
  $v_z(Y_z)=v_{0z}(Y_z)+{\alpha_z}\:v_{1z}(Y_z)$.
To zeroth order, we have the usual linear equation:
\begin{equation}
\frac{\partial^2 v_{0z}}{\partial Y_z^2}-v_{0z}=0,
\label{leq}
\end{equation}
with $v_{0z}$ satisfying the boundary conditions  (\ref{bcds}), (\ref{bc}). 
The solution is:
\begin{equation}
v_{0z}(Y_z)=c_z \sinh(Y_z),
\label{linsol}
\end{equation}
where
\begin{equation}
\label{adef}
c_{z}=\frac{{e} \lambda_z H_a\sin\psi}{c\cosh(Y_{z s})}.
\end{equation} 
The nonlinear part $v_{1z}$ satisfies:
\begin{equation}
\frac{\partial^2 v_{1z}}{\partial Y_z^2}-v_{1z}
+v_{0z}^3=0.
\label{vsin}
\end{equation}
The boundary conditions are
%\begin{equation}
$\left. {\partial v_{1z}}/{\partial Y_z} \right |_{Y=Y_{z s}}=0$ 
%\end{equation}
and 
$v_{1z}(0)= 0$.
The complete solution to Eq. (\ref{vsin}) is 
found by elementary methods and is given by:
\begin{equation}
v_{1z}(Y_z)=({1}/{8})c_z^3 \left[{c}_1 \sinh(Y_z)+3 Y_z\cosh(Y_z)-({1}/{4})
 \sinh(3Y_z)\right], 
\label{v1sol}
\end{equation}
where
%\begin{equation}
$c_{1}=(3/2)(\sinh(2 Y_{z s})-2 Y_{z s})\tanh(Y_{z s})-9/4$. 
%\end{equation}

The magnetic field in the sample can be calculated from the field 
${\bf v}$ via Eq. (\ref{londoneq}). Including also the purely linear
component arising from  $v_x$ we obtain:
\begin{mathletters}
\label{hsol}
\begin{equation}
H_x(Y_z)=\frac{H_a 
\sin\psi}{\cosh(Y_{z s})}\left[\cosh(Y_z)+\frac{1}{8}\left(\frac{H_a\sin\psi 
}{H_{0 z}\cosh(Y_{z s})}\right)^2 f_H(Y_z)\right],
\label{fieldsa}
\end{equation}
\begin{equation}
H_z(Y_x)=\frac{H_a\cos\psi}{\cosh (Y_{x s})}\cosh(Y_x),
\end{equation}
where the nonlinear depth dependence is 
contained in $f_H$:
\begin{equation}
f_H(Y_z) = 3 Y_z \sinh(Y_z)-(3/4)\cosh(3 Y_z)+(c_1+3)\cosh(Y_z),
\end{equation}
\end{mathletters}
and we have introduced the characteristic field
\begin{equation} 
H_{0 i}=\frac{\phi_0}{\pi ^2 \lambda_i \xi_i}.
\end{equation}
Here $\phi_0$ is the superconducting
flux quantum, and $\xi_i={v_{f i}}/({\pi\Delta_0})$ is the local 
coherence length.  
As opposed to the
d-wave case where the nodal lines give rise
to a quadratic nonlinear contribution, we now find a nonlinear effect
cubic in the applied field. Physically, this is quite transparent: the
phase space volume available to the quasiparticle excitations increases
as the cube of the field for point nodes, and as the square for line
nodes.
The nonlinear term anisotropically increases the magnetic field penetration 
because of 
quasiparticle occupation near the nodes, i.e., fewer Cooper pairs are 
participating in the current responsible for bulk flux exclusion. 

We can gain 
some insight into these results
by examining the spatial dependence of the nonlinear 
part of the field as displayed in Fig. \ref{jfig}.
There the quantity $H_{nlx}$, defined as the last term in 
(\ref{fieldsa}) normalized to unity at its
maximum, is 
plotted as a function of dimensionless distance $D$
from the surface ($D=Y_{sz}-Y_z$). The thickness of the sample is 
taken to be $d >>\lambda_z$ so that 
the behavior shown is that corresponding to a thick slab.
The 
nonlinear field is constrained by the boundary conditions to vanish at 
the surface: the boundary condition implies an extremum for  
the nonlinear flow field at the surface and since
$H_x \propto \partial{v_{1z}}/\partial{Y_z}$, we see that
$H_{nlx}$ must vanish there. It then increases  rapidly 
reaching its maximum at about one half of a penetration depth and then 
decays exponentially inside the sample, as does the linear part. 
 Thus arises
the characteristic maximum of the nonlinear field seen in this Figure.

The current is most easily obtained from ${\bf H}$ through
Amp\`ere's law for steady-state currents, 
which gives the result:
\begin{mathletters}
\label{jsol}
\begin{equation}
j_z(Y_z)=-\frac{c H_a \sin\psi}{4 \pi 
\lambda_z\cosh(Y_{z s})}\left[\sinh(Y_z)-\frac{1}{8}\left(\frac{H_a\sin\psi 
}{H_{0 z}\cosh(Y_{z s})}\right)^2 f_J (Y_z)\right],
\label{jsola}
\end{equation}
\begin{equation}
j_x(Y_x)=\frac{c H_a\cos\psi}{4 \pi \lambda_x\cosh (Y_{x s})}\sinh(Y_x).
\end{equation}
The $Y_z$ dependence of the nonlinear current is contained in $f_J (Y_z)$ 
which is 
given by:
\begin{equation}
f_J (Y_z)= -3 Y_z \cosh(Y_z)+(9/4)\sinh(3 Y_z)-(c_1+6)\sinh(Y_z).
\end{equation}
\end{mathletters}
In Eq. (\ref{jsola}), the first term on the right is the linear contribution, 
proportional to the applied field, while the second term is
the nonlinear correction, modifying the 
total current. 

We now turn to the case where the OP is of the form (\ref{deltacos}). For
simplicity, we consider tetragonal symmetry and quote the
results for the orthorhombic case later. 
The current in Eq. 
(\ref{jcos}) can be simplified using that the magnetic field is applied 
in the $x-z$ plane, so that $v_y$ is zero. The current is 
substituted 
in (\ref{lap}) to give, for the component containing the nonlinear term:
\begin{equation}
\frac{\partial^2 v_x}{\partial Y_x^2}-v_x
+\frac{\left |v_x \right |}{v_{c x}}v_x=0. 
\label{ucos}
\end{equation}
The method of solution is identical to that above with the only major 
difference being that the expansion parameter is now
$\alpha_x =({v_{f x}}/{\Delta_0})$, linear rather than quadratic. 
The linear velocity field $v_{0x}$ is  
$v_{0x} = -c_x \sinh(Y)$, while the nonlinear term is written as
\begin{equation}
v_{1x}(Y_x)=(1/6)c_x |c_x|\left[\cosh(2 
Y_x)-4\cosh(Y_x)+4 c_2\sinh(Y_x)+3\right],
\label{vxcos}
\end{equation}
where
\begin{equation} 
c_{x}= \frac{{e} \lambda_x H_a\cos\psi}{c\cosh(Y_{x s})}, \qquad
c_2=\tanh(Y_{x s})-\sinh(Y_{x s}).
\end{equation}

The magnetic field  is calculated again via the London
equation. In this case the nonlinear part is (as in the d-wave case)
proportional to $H_a^2$, 
rather than to $H_a^3$ as was found for the point nodes. 
This follows again from phase-space arguments. After including  the
contribution from the purely linear component of ${\bf v}$ one finds:
\begin{mathletters}
\label{hsol2}
\begin{equation}
H_z(Y_x)=\frac{H_a 
\cos\psi}{\cosh(Y_{x s})}\left[\cosh(Y_x)+\frac{1}{6}\left(\frac{H_a 
|\cos\psi|}{H_{0 x}\cosh(Y_{x s})}\right) g_H(Y_x)\right], 
\label{fieldca}
\end{equation}
\begin{equation}
H_x(Y_z)=\frac{H_a\sin\psi}{\cosh (Y_{z s})}\cosh(Y_z).
\end{equation}
Here $g_H$, which determines the nonlinear contribution
 to the magnetic field penetration, is given by:
\begin{equation}
g_H(Y_x) = -2 \sinh(2 Y_x)+4 \sinh(Y_x)-4 c_2\cosh(Y_x).
\end{equation}
\end{mathletters}

It is again useful to plot the nonlinear component of the magnetic field.
We consider $H_{nlz}$, the last term in (\ref{fieldca}) normalized
to its maximum value.
Fig. \ref{jfig2} shows $H_{nlz}$ plotted 
versus dimensionless distance $D$ from the
surface ($D \equiv Y_{sx}-Y_x$). 
 One sees again the rapid increase of the 
field near the surface of the sample, followed by the usual
exponential decay. The plot is very similar to that in Fig. \ref{jfig}
for point nodes, but the field decays less 
rapidly into the sample. Within about three penetration depths, 
the nonlinear field is reduced to 20\% of its maximum magnitude.

The total current is composed of linear and nonlinear terms, as
found from Amp\`{e}re's law:
\begin{mathletters}
\label{jsol2}
\begin{eqnarray}
j_x(Y_x)&=&\frac{c H_a \cos\psi}{4 \pi 
\lambda_x\cosh(Y_{x s})}\left[\sinh(Y_x)-\frac{1}{6}\left(\frac{H_a 
|\cos\psi|}{H_{0 x}\cosh(Y_{x s})}\right) g_J (Y_x)\right], \\
j_z(Y_z)&=&-\frac{c H_a\sin\psi}{4 \pi \lambda_z\cosh (Y_{z s})}\sinh(Y_z).
\end{eqnarray}
\end{mathletters}
where 
\begin{equation}
g_J (Y_x) = 4[ \cosh (2 Y_x) -  \cosh (Y_x) +  c_2 \sinh (Y_x)],
\end{equation}
is the function determining the penetration of the nonlinear
currents.

We can now use our solutions for both OP's to derive expressions for the 
experimentally relevant quantities. 

\subsection{The transverse magnetic moment}
The expression for the magnetic moment in terms of the currents
 is:\cite{jackson}
\begin{equation}
{\bf m}= \frac{1}{2 c} \int 
d {\bf r} {\bf r} \times {\bf j(v)}.
\label{mom}
\end{equation}
By making use of standard identities and the parity of ${\bf v}$,
Eq. (\ref{mom})
can be rewritten\cite{zv2,brandt} more conveniently as:
\begin{equation}
m_{x,z}=-\frac{V H_{a\:x,z} } {4 \pi} \mp 
\left. \frac{A c \:v_{z,x}}{2 \pi e}\right|_{y=\frac{d}{2}},
\label{msur}
\end{equation}
where ${A}$ is the surface 
area of the plane along which the field
is applied, ${V}$ is the volume of the sample, and
we have used that ${\bf v}$ is odd in $z$. 
The magnetic moment perpendicular to the applied field is given by 
$m_{x} \cos \psi -m_{z} \sin \psi$.  The linear terms of the velocity 
fields contribute to this quantity only if there is anisotropy in
the penetration
depth tensor. 
In that  case, the linear term in the transverse
magnetic moment, denoted by
$\tilde{m}_{\bot}$, is:
\begin{equation}
\tilde{m}_{\bot}=\frac{1}{4\pi}A H_a (\lambda_z-\lambda_x)\sin 2\psi.
\end{equation}
This term can be distinguished from the nonlinear contribution, 
$m_\perp$, because
of its different field and angular dependences.

For  an OP with  point nodes, (\ref{deltasin}),
 ${m}_{\bot}$
is obtained from (\ref{msur}), (\ref{v1sol}) and (\ref{hsol})
as:
\begin{equation}
m_{\bot}(\psi)=\frac{1}{4 \pi} A \lambda_z H_a  
\left(\frac{H_a}{H_{0 z}}\right)^2[(3^{3/2}/32) f_s(\psi)]{\cal K_{S}}(Y_{z s}).
\label{msin}
\end{equation}
This quantity is proportional to 
the cube of the applied field, rather than, as in the d-wave
case,\cite{ys,sv,zv1,zv2}
to the square. This reflects the reduced phase space when 
the nodal regions are points rather than lines on the FS. Thus larger
values of $H_a$ are very advantegeous provided that $H_a$ is kept
below the field of first flux penetration,
$H_{f1}$, so that the sample remains in the Meissner regime.
The angular dependence of $m_{\bot}(\psi)$ is contained in $f_s(\psi)$ 
which is normalized to unity at its maximum and given by:
\begin{equation}
\label{fsin}
f_s(\psi) = (16/3^{3/2}) \cos \psi \sin^3 \psi, \qquad
\end{equation}
while the dependence of $m_{\bot}$ on the material thickness $d$
is given by the function
${\cal K_{S}}$,
\begin{equation}
{\cal K_S}\left(Y_{z s}\right)= (1/2){\mathrm{sech}}^4(Y_{z s})\left[3 Y_{z s} - 
2\sinh(2 Y_{z s}) + (1/4) \sinh(4 Y_{z s})\right]. 
\label{Ksin}
\end{equation}
The torque associated with $m_\perp$ is simply obtained by multiplying
(\ref{msin}) by $H_a$, therefore it has the same thickness and 
angular dependence.

The function $f_s(\psi)$ is displayed as the solid line in  Fig. \ref{figMM}.
The transverse magnetic moment and torque in this case are maximal for the 
field direction corresponding to
$\psi = \pi/3$, and vanish at directions corresponding to
the nodes or antinodes of the OP. 
The $\pi$ periodicity of $m_{\bot}$ matches that of the energy since the 
angular dependence of the quasiparticle
energy arises solely from that of $|\Delta(\theta)|^2$.

Since $m_\perp$ is an extensive quantity and it is often the
case that larger samples can be made in film form rather than grown
as free standing crystals, it is of considerable interest to
examine the thickness dependence of the results, as given by ${\cal K_S}$.
The behavior of ${\cal K_S}$ is displayed 
in Fig. \ref{fig3}, where 
${\cal K_S}$ is plotted (solid line)
 as a function of $Y_{sz}\equiv d/(2\lambda_z)$.
It is seen that ${\cal K_S}$ increases rapidly with $Y_{sz}$, reaching 
$90\%$ of its 
maximum value of unity when $d \simeq 5 \lambda$.
If the sample  is a thick slab, $d\gg\lambda_z$, then
${\cal K_S}\rightarrow 1$, so that $m_{\bot}$ is (for
the same area $A$) maximal and 
independent of $d$. 
 On the other hand, the decrease of ${\cal K_S}$ with
 thickness is substantial: for films where $d = \lambda_z$, ${\cal K_S}$ 
is  ${\cal K_S}(1/2) = .017$, a reduction
of over 80\%. Such a decrease, however, may very
well be compensated
by a larger increase in $A$, compared to a free
standing crystal. However, for extremely thin films,
$d\ll \lambda_z$, then ${\cal K_S} \simeq 1/40(d/\lambda_z)^5$,
and despite the increase of $H_{f1}$ in thin films, the amplitude of the
signal would almost certainly be too small\cite{andre}.

When the order
parameter is of the form (\ref{deltacos}),
the results obtained from the previous expressions (\ref{vxcos}) and
(\ref{hsol2}) for
a compound with tetragonal symmetry yield, when substituted
in (\ref{msur}), 
the expression:
\begin{equation}
m_{\bot}(\psi)=\frac{1}{4 \pi} A {\lambda_{x}} H_a \left 
(\frac{H_a}{H_{0 x}}\right)[(4/3^{5/2}) f_c(\psi)]{\cal K_{C}}(Y_{x s}).
\label{mcos}
\end{equation}
We have introduced
the fuctions
$f_c(\psi)$, and  $\cal K_C$ characterizing, respectively,
the angular and the thickness dependences of the result.
They are given by:
\begin{mathletters}
\label{fcos}
\begin{eqnarray}
f_c(\psi)& =& (3^{3/2}/2)|\cos \psi | \cos \psi \sin \psi, \\
{\cal K_C}\left(Y_{x s}\right)&=& \left[{\mathrm{sech}}(Y_{x s}) - 1 
\right]^2\left[1+2 \: {\mathrm{sech}} (Y_{x s})\right]. 
\end{eqnarray}
\end{mathletters}
The nonlinear transverse moment is now proportional to the square
of the applied field, as in the d-wave case, because of the linear
character of the nodal regions.
The function $f_c(\psi)$ is normalized to unity and it is plotted
as the dashed line of Fig. \ref{figMM}.
It is seen there that the angular signature is
different from that in the previous case: 
$m_{\bot}$ has now a maximum when $\psi = \arctan(\sqrt{2}/2)$, 
although it is zero again for fields applied along
the nodes and antinodes ($\psi=\pi/2,0$) in the OP. As in the
previous case, ${\cal K_C}$ is small for small thickness. 
If we are 
dealing with a thick slab, ${\cal K_C}\rightarrow 1$, 
but when $d \sim \lambda_x, \:{\cal 
K_C} \approx .036$. In the limit $d \ll \lambda_x$, 
 ${\cal K_C} \simeq 3/64(d/\lambda_x)^4$.
 As seen in Fig. \ref{fig3}, the overall characteristics of 
 ${\cal K_C}$ (dashed line) are very similar to those of ${\cal K_S}$, 
 but ${\cal K_C}$ is larger in magnitude throughout. Both 
 curves show a 50\% drop in signal when $d\simeq 3 \lambda_x$.

 It is somewhat tedious but straightforward to generalize, 
 starting from the form (\ref{jcos}), the
 calculation of the fields and currents for this order
 parameter to the case where there is penetration depth anisotropy
 in the $a-b$ plane. 
  The result is that, 
 when the sample is rotated about the $b$ axis, 
 Eq. (\ref{mcos}) is simply modified to:
\begin{equation}
m_{\bot}(\psi)=\frac{1}{4 \pi} A \lambda_x H_a \left 
(\frac{H_a}{H_{0 x}}\right)[(4/3^{5/2}) f_c(\psi)]{\cal 
K_{C}}(Y_{x s}){\Gamma}_x,
\label{mcosa}
\end{equation}
where  ${\Gamma}_x \equiv (1/5)(4+
\delta^2)$. 
Thus, the angular, field and thickness dependences remain the same, while
only an overall anisotropy factor is needed.

\subsection{Penetration depth}

Measuring the field dependence of the penetration depth at low temperatures
is another possible way of exploring the nonlinear Meissner effect.
The reduction of the current via 
quasiparticle population results in a lower superfluid density and hence 
a larger 
penetration depth.
 Indeed,
this was the first quantity  studied\cite{ys} in this area, although
it is only very recently that  
experimental measurements\cite{bid} have been attempted for HTSC's. 

In the presence of nonlinear effects, several possible definitions of the
penetration length which coincide in the linear limit give slightly
different results. The appropriate definition depends on the experimental
setup. We have calculated above the 
spatial current and field distributions, and these results can be used
to obtain the nonlinear contributions to $\lambda$ for any definition.
We briefly illustrate this here by computing 
the components of the penetration depth along the $x$ and $z$ directions 
via the definition:
\begin{equation}
\label{pen}
\frac{1}{\lambda_{z}({\bf H}_a)}=\frac{1}{ \lambda_z H_a}\left (\frac{\partial 
H_x}{\partial Y_z}\right)_{Y=Y_{z s}}, 
\qquad
\frac{1}{\lambda_{x}({\bf H}_a)}=\frac{1}{ \lambda_x H_a}\left (\frac{\partial 
H_z}{\partial Y_x}\right)_{Y=Y_{x s}},
\end{equation}
where $\lambda_i$ is the zero field penetration depth along the $i$ direction.
We will assume that the 
field is applied along a 
symmetry direction for each OP and will not 
place any restrictions on the thickness $d$.

We consider first the order parameter with point nodes, (\ref{deltasin}).
The penetration depth when the applied field is perpendicular
to the z-axis,  ($\psi = \pi/2$) is obtained from (\ref{pen}) and
(\ref{hsol}):
\begin{equation}
\label{pens}
\frac{1}{\lambda_z ({H_a})}=\frac{1}{\lambda_z} \left \lbrace \tanh(Y_{z s})- 
\frac{3}{32}\left(\frac{H_a}{H_{0z}}\right)^2 {\cal L_S}\right \rbrace,
\end{equation}
where 
${\cal L_S}={\mathrm{sech}}^4 (Y_{z s}) [-4 Y_{z s} +\sinh(4 Y_{z s})]$.
The most obvious difference between this and the results for d-wave is that
the field correction is proportional to the square of the field, rather
than to the field itself. This follows, once more, from phase space arguments.
For a thick slab, ${\cal L_S}\rightarrow 8$, while
in the very
thin film limit, ${\cal L_S} \approx ({4}/{3}) 
({d}/{\lambda_z})^3$.

For the OP given in (\ref{deltacos}) one similarly gets:
\begin{equation}
\label{penc}
\frac{1}{\lambda_x ({H_a})}=\frac{1}{\lambda_x} \left \lbrace \tanh(Y_{x s})- 
\frac{2}{3}\left(\frac{H_a}{H_{0x}}\right){\cal L_C} \right \rbrace,
\end{equation}
where ${\cal L_C}=1 - {\mathrm{sech}}^3(Y_{x s})$, which goes to unity
for a thick slab.  For a 
very thin film superconductor, 
${\cal L_C}={2}/{3} ({d}/{\lambda_x})^2$.
Here the penetration depth correction, 
${\delta\lambda}/{\lambda_x}$ is linear in the applied filed
as in the d-wave case, as a result of the presence in both cases of
nodal lines. 
The dependence on thickness in the very thin film limit is cubic in 
$d$ for the point nodes and quadratic for line nodes.

\section{Discussion}
\label{disc}
 
We have calculated the magnitude of the nonlinear
electrodynamics effects as a function of field and angle.  We
believe that some remarks are now in order as to the feasibility of observing
effects of the rough magnitude of those predicted. In making these
remarks, we do not claim any experimental expertise
in the relevant areas. We have in
mind compounds with materials characteristics such as those of
$\rm{U B e_{13}}$ or ${\rm Sr_2RuO_4}$. In mentioning these
compounds, we do not intend
to propose that any of them belong to
a  specific pairing state. 
We merely wish to roughly estimate the level of the
signal that  would be predicted
in the event that the material
turned out to have a pairing state with a certain nodal structure.
Our
considerations can straightforwardly be extended to other materials.

First, our calculations have been performed, strictly speaking, in the low
temperature limit.  In practice, this means\cite{sv,ys1} the temperature
regime in the region
$T<\tilde{T}(H)\lesssim \Delta_0 H/H_0$, so that thermal excitations do not 
destroy\cite{sv} the nonlinear Meissner effect. Assuming that,
in order to maximize the signal, the applied field is 
close
to $H_{f1}$, the field of first flux penetration, 
this still implies that the experiments must be performed at
temperatures well below $T_c$. This means, for the materials of interest, at
dilution refrigerator temperatures.  This is undeniably a disadvantage when
compared with the situation for HTSC's, but one that can be overcome by using
torque magnetometry\cite{tm} or
torsion oscillator\cite{to} techniques to measure the 
torque associated with the
transverse moment.  These techniques can be
adapted to use in conjunction with
a dilution refrigerator and their sensitivity can
surpass\cite{allen} that of the
SQUID methods used for HTSC's\cite{bu,bhat}. These considerations pertain to the transverse 
moment and the
associated torque. Several groups are planning, in the
context of checking the very low temperature behavior of $\lambda$
in HTSC's for deviations from the linear power law behavior, measurements of
$\lambda$ at dilution refrigerator temperatures. These techniques
could be combined
with the high resolution methods already employed\cite{bid} to measure the field
dependence of $\lambda$.

We consider next the magnitude of the low $T$ effect.  The maximum amplitude of
the transverse moment depends on the values of the penetration depth, 
the characteristic field $H_{0i}$,
and  the maximum field one can apply while remaining in the Meissner regime,
which is $H_{f1}$.  Because $m_\perp$ is an extensive quantity, it
also depends on the size (specifically the surface area) of the samples
available.  As an illustrative exercise, we have estimated a putative
signal for the 
transverse magnetic moment amplitude
for a number of  compounds by getting from the 
literature\cite{ruth,hf,bhat,gml,pam,rnk} values of available crystal sizes and 
of the experimental
parameters (such as penetration depths, and correlation lengths
in the appropriate directions) 
that appear in our  expressions. 
These values are subject to very considerable
uncertainty and in most cases different references do not agree with each other,
but they are sufficient for the purposes of our
exercise.  By inserting them in the appropriate
formulae, (\ref{msin}), and (\ref{mcos}), we obtain numerical values of the 
possible signal.  The
results  are summarized in Table I, where we present our
estimate for the maximum amplitude $M_\perp$ for several materials.
$M_\perp$ is defined as the
value of $m_\perp$ for a thick slab at $H_a=H_{c1}$ and at the angle
$\psi$ for which $m_\perp$ is maximal. The critical field  $H_{c1}$ is 
calculated from\cite{dg}  
$H_{c1}=({\phi_0}/({4\pi\lambda^2}))\ln({\lambda}/{\xi})$ and is 
used as a conservative approximation to $H_{f1}$ since $H_{c1}$ is  
smaller (by a large factor\cite{com} for YBCO) than $H_{f1}$.
We have also included in the Table, for the purposes of comparison,
one typical HTSC compound (YBCO), with the signal in that case computed
from Ref.\onlinecite{zv2} for a d-wave state.
For the other materials, 
we have assumed the OP in Eq. (\ref{deltasin}) for 
${\rm Sr_2RuO_4}$, while for the listed heavy fermion materials 
and organic salt we 
have taken the OP  of 
Eq. (\ref{deltacos}).  It bears repeating
that  these
choices are illustrative and  do not imply any judgement
 on our part as to the
likelihood of what the pairing state
actually might be. Rather, our point is that 
the  techniques in this paper can
 be implemented to infer the nodal structure of the pairing state.
The results in Table I are expressed in physical units and also, for
purposes of comparison, as ratios to the corresponding estimate for 
YBCO in a d-wave state.  The
numbers in the table are very encouraging:  they are in all cases 
comparable to or
larger than those for YBCO and always comfortably exceed the resolution of
the experimental techniques discussed above.\cite{to,tm} For the penetration 
depth
results, the situation is similarly favorable, since the changes in the
penetration depth induced by a field close to $H_{f1}$ are considerably larger
than the lower limit (a few ${\rm \AA}$ resolution)
 already achieved\cite{bid} in
YBCO.

We have to consider also the limitations of this work and the presence of other
phenomena, besides temperature excitations, that may reduce the signal.  First,
there is the question of impurities.  As has been seen in the context of
YBCO,\cite{sv,ys1} good quality samples characterized by a transition
temperature not appreciably degraded, and by the appropriate power law 
behavior of $\lambda$ with temperature, should exhibit a signal
substantially
of the magnitude calculated here for a clean system.  The decrease in the
nonlinear signal associated with nonlocal effects at lower fields can also
complicate the situation\cite{li1}.  However, these effects are quite small for
fields close to $H_{f1}$ in the typical situation where this field\cite{com} is
considerably larger than the equilibrium $H_{c1}$.  In any case nonlocal effects
are absent for several special crystal orientations\cite{leg,li2}, which can
then be chosen.  There also important questions as to what the effect of using
more realistic forms (still containing point or line nodes) of the order
parameter would be, or of including in more detail the local value properties
 of the Fermi
surface.  All told, however, we believe that the estimates in
the Table show that there is a sufficient cushion between the maximum value and
the experimental resolution so that one can expect an observable signal.

In conclusion, we have calculated here the  nonlinear signal arising
from the presence of point or line nodes in some simple p-wave order parameters.
We have shown that there is a likelihood that these effects will be observable
in materials currently being studied.  The results given can straightforwardly
be extended, if and when the experimental situation warrants it, to
the study of the low frequency response\cite{lfr},
to more
complicated or mixed order parameters, and a more
general  node spectroscopy procedure for
p-wave materials can be performed as in Ref.\onlinecite{zv2}.

\acknowledgments
We are indebted to Igor \u{Z}uti\'{c} for conversations on the theories presented
in this paper and for reading a draft of the manuscript.
We thank Paul Crowell and Allen Goldman for numerous conversations
concerning what can and cannot be done experimentally. This work was
supported in part by the Petroleum Research Fund, administered by the ACS.

\begin{figure} 
\caption{ Spatial behavior of the nonlinear field when the OP has
point nodes. The quantity ploted is $H_{nlx}$, 
as given by the last term in (\ref{fieldsa}), normalized to unity at its 
maximum. This quantity is
plotted as a
function of  $D$,  the distance from the surface in units of
$\lambda_z$: $D \equiv Y_{s z}-Y_z$. The sample has thickness $d>>\lambda_z$.
The nonlinear field increases rapidly until the distance from the surface 
is about half
$\lambda_z$ and then decreases exponentially, as in the 
usual linear case. }  
\label{jfig} 
\end{figure}

\begin{figure} 
\caption{The nonlinear $z$ component of the 
magnetic field for a material with an OP of the form (\ref{deltacos}).
The quantity plotted is $H_{nlz}$, the last term in (\ref{fieldca}),
normalized to unity at its maximum. It is
plotted versus dimensionless distance from the 
surface: $D\equiv Y_{sx}-Y_x$. The material thickness is $d >> \lambda_x$. 
The behavior is qualitatively similar to that shown in the previous Figure,
but the maximum occurs at a somewhat greater depth.}  
\label{jfig2} 
\end{figure}

\begin{figure} 
\caption{The angular dependence of the normalized transverse magnetic
 moment is plotted versus $\psi$ (the angle between ${\bf H}_a$ and
the $z$ axis, in radians). The solid line is the function $f_S(\psi)$ given in
(\ref{fsin}) for an OP of the form (\ref{deltasin}), while the dashed line is
$f_C(\psi)$ from
(\ref{fcos}) for the OP (\ref{deltacos}). Both functions
 are normalized to their maximum values. Their maxima are at $\psi=\pi/3$
 and $\psi=\arctan(\sqrt{2}/2)$ respectively and their periodicity is $\pi$.}  
\label{figMM} 
\end{figure}

\begin{figure} 
\caption{The symbol {\rm K} stands for the functions ${\cal 
K_{S}}$ (solid curve) and ${\cal 
K_{C}}$ (dashed curve), which characterize the thickness
dependence of $m_{\bot}$. They are plotted versus the dimensionless
thickness $Y_s$ ($Y_s$
represents ${d}/{2\lambda_z}$ for ${\cal K_{S}}$ and  
${d}/{2\lambda_x}$ for the ${\cal K_{C}}$ plot.)
 The exponential increase of the functions towards unity is rapid and the 
 corresponding bulk regime arises when the material thickness $d$ is 
about five penetration depths.}  
\label{fig3} 
\end{figure}

%\begin{table}
%\caption{Spin singlet pairing states in a $CuO_2$ plane.}
%\label{table1}
%\begin{tabular}{ccc}
%Symmetry Class&Informal name&Representative State\\
%\tableline
%$A_{1g}$ &$s$ $(s^+)$    &$const.$\\
%$A_{2g}$ &$g$ $(s^-)$ &$xy(x^2-y^2)$\\
%$B_{1g}$ &$d_{x^2-y^2} $ &$x^2-y^2$\\
%$B_{2g}$ &$d_{xy}$       &$xy$\\
%\end{tabular}
%\end{table}
\begin{table}
\caption{Illustrative estimates of possible nonlinear
signal for various materials of uncertain OP nodal structure. 
Values for d-wave YBCO are also given for comparison purposes.
The magnetic 
field is in Gauss, and the maximum transverse magnetization
amplitude, $M_{\bot}$, as defined in the text, is given in 
($\times 10^{-8}$) emu's. 
The quantities $\lambda$ and $H_0$ are for the relevant directions
(see text).  }
\label{table1}
\begin{tabular}{ccccccc}
Compound&Area $(mm^2)$&$\lambda$ (\AA)&$H_0 $&$\frac{H_{c1}}{H_0}$&$M_\bot 
$&$\frac{M_{\bot}}{M_{\bot}(YBCO)}$ \\
\tableline
${\rm Y B C O}$
%\tablenote{For ${\rm Y B C O}$ we use the expression 
%for $m_{\bot}$ found in Ref.\onlinecite{zv1}.}
&1.8&1400&7607&.048&2.1&1\\
\tableline
${\rm U B e_{13}}$&5&11000&204&.032&2.4&1.1\\
\tableline
${\rm U P t_3}$&150&15000&71&.045&66&31.4\\
\tableline
${\rm Sr_2RuO_4}$&25&1940&147&.29&22.5&10.7\\
\tableline
${\rm (TMTSF)_2ClO_4}$&1&5000&43&.25&2.7&1.3 \\
\end{tabular}
\end{table}

\begin{references}
\bibitem{below} As explained below, any contributions 
associated with internal degrees of freedom of the OP should
be unimportant in solid systems at small applied fields.
\bibitem{par} Possible contributions arising from  finite spin of
the Cooper pairs are negligible compared to the diamagnetic terms.
\bibitem{rotate} For any situation where the crystallographic
direction of the point nodes, or the plane of the line
nodes, is different from that assumed here, our considerations and
results apply after performing the appropriate change of labels.
\end{references}
\end{document}